\documentclass[10pt,a4paper]{article}
\date{}
\title{Partition Function of the Harmonic Oscillator on a Noncommutative Plane}
\author{I. Jabbari, A. Jahan, Z. Riazi\\
Department of Physics, Amirkabir University of Technology\\
 P. O. Box 15875-4413, Tehran, Iran\\
jahan@riaam.ac.ir}
\usepackage[english]{babel}
\begin{document}
\maketitle
\begin{abstract}
We derive the partition function of the one-body and two-body systems of classical noncommutative harmonic oscillator in two dimensions. Then, we employ the path integral approach to the quantum noncommutative
harmonic oscillator and derive the partition function of the both systems at finite
temperature.\\
\end{abstract}
\section{Introduction}
It seems natural to explore the noncommutative (NC) version of quantum mechanics as the one-particle low energy sector of the noncommutative quantum field theory, which recently has gained a great interest mainly due to the arguments based on the String/M theories [1-5]. The following set of relations among the coordinates and momenta (setting $\hbar=1$)
{\setlength\arraycolsep{2pt}
\begin{eqnarray}\label{1}
[\hat{x}_{i},\hat{x}_{j}]&=&i\theta_{ij},\\
{}[\hat{x}_{i},\hat{p}_{j}]&=&i\delta_{ij},\\
{}[\hat{p}_{i},\hat{p}_{j}]&=&0,
\end{eqnarray}}
will characterize the noncommutative version of the two dimensional quantum mechanics, in which $\theta_{ij}$ stand for the constant antisymmetric matrix. However, one may decide to maintain the usual canonical structure of the quantum mechanics and implement the noncommutativity through defining a new set of commutating coordinates $x_{i}$ via [6-11]
\begin{equation}\label{2}
x_{i}=\hat{x}_{i}+\frac{1}{2}\theta_{ij}p_{j}.
\end{equation}
where the summation is implied over the repeated indices. Therefore, for the
interaction potential $\hat{U}(\hat{x}_{i})$ defined on a noncommutative plane, one gets the effective $\theta$-dependent potential
in usual commutative plane as
\begin{equation}\label{3}
U_{\theta}(x_{i},\bar{p}_{i})=U_{\theta}(x_{i}-\theta_{ij}p_{j}/2),\quad\bar{p}_{i}=\theta_{ij}p_{j}.
\end{equation}
On the other hand, path integral approach to the noncommutative quantum mechanics has revealed in series of
the recent works [12-14]. It is demonstrated by the authors that how the noncommutativty of
space could be implemented in Lagrangian formulation of the quantum mechanics. In this letter, we first derive the partition function of a two dimensional classical noncommutative harmonic at finite temperature. We also consider a two-body system of particles bounded by a harmonic oscillator potential.
Then, we focus on the quantum one-body and two-body problem of noncommutative harmonic oscillator and employ the formalism developed in [12-14] to incorporate the effect of noncommutativty of space on the thermodynamics of a two dimensional harmonic oscillator. It is
shown that the result is in accordance with one obtained via the method based on the Hamiltonian approach to
the noncommutative harmonic oscillator.
\section{Classical NC Harmonic Oscillator}
The Hamiltonian governing the dynamics of a classical harmonic oscillator on a noncommutative plane is
\begin{equation}\label{4}
\hat{H}=\frac{1}{2m}(\hat{p}^{2}_{1}+\hat{p}^{2}_{2})+\frac{1}{2}m\omega^{2}(\hat{x}^{2}_{1}+\hat{x}^{2}_{2})
\end{equation}
On implementing the transformation (4), one finds the $\theta$-dependent Hamiltonian in usual
commutative space as
\begin{equation}\label{5}
H_{\theta}=\frac{\kappa}{2m}\vec{p}^{\,2}+\frac{1}{2}m\omega^{2}\vec{x}^{\,2}
+\frac{1}{2}m\omega^{2}\vec{\theta}\cdot{}\vec{x}\times\vec{p}.
\end{equation}
with $\kappa=1+\frac{1}{4}m^{2}\omega^{2}\theta^{2}$ and $\vec{\theta}=(0,0,\theta)$. Thus, the finite temperature partition function reads
{\setlength\arraycolsep{2pt}
\begin{eqnarray}\label{6}
Z_{\theta}(\beta)&=&\int_{-\infty}^{+\infty}{d^2{p}}\int_{-\infty}^{+\infty}{d^2{x}}e^{-\beta{H_{\theta}(\vec{p},\vec{x})}},\\\nonumber
&=&\int_{-\infty}^{+\infty}{d^2{p}}\,e^{-\frac{\beta\kappa}{2m}{\vec{p}^{2}}}
\int_{-\infty}^{+\infty}{d^2{x}}\,e^{-\frac{1}{2}m\beta\omega^{2}{\vec{x}^{2}}}
e^{-\frac{1}{2}m\omega^{2}\vec{\theta}\cdot{}\vec{x}\times\vec{p}},\\\nonumber
&=&\bigg(\frac{2\pi}{\beta\omega}\bigg)^{2}.
\end{eqnarray}}
Of course, this is the partition function of a usual 2D oscillator. Thus, there is no contribution arising from the noncommutativity of the plane on the partition function.
\section{Quantum NC Harmonic Oscillator}
In this section (mainly taken from [15]) we derive the partition function of a quantum noncommutative harmonic oscillator by means of the results derived earlier in [12-14]. The corresponding Lagrangian of the $\theta$-dependent Hamiltonian (7) can be achieved by means of the Legender transformation $
L_{\theta}(\dot{\vec{x}},\vec{x})=\vec{p}\cdot\dot{\vec{x}}-H_{\theta}(\vec{x},\vec{p})$, upon replacing for the momenta from
$\dot{\vec{x}}=\partial{H_{\theta}}/\partial{\vec{p}}$. Hence, the Lagrangian associated with the Hamiltonian (7) will be [12-14]
\begin{equation}\label{7}
L_{\theta}=\frac{m}{2\kappa}(\dot{x}^{2}_{2}+\dot{x}^{2}_{1})-\frac{m}{2\kappa}\omega^{2}
(x^{2}_{2}+x^{2}_{1})+\frac{\theta}{2\kappa}{m^{2}}\omega^{2}(\dot{x}_{2}x_{1}-\dot{x}_{1}x_{2}).
\end{equation}
The above Lagrangian admits the equations of motion as differential equations of rank four.
When the solutions are inserted in (9) and integrated over the time variable $\tau$, one
finds for the action
{\setlength\arraycolsep{2pt}
\begin{eqnarray}\label{8}
S_{\theta}(\vec{x}^{\,\prime\prime},\vec{x}^{\,\prime})
&=&\frac{m\omega}{2\sqrt{\kappa}\sin(\omega\tau\sqrt\kappa)}\bigg[(\vec{x}^{\,\prime{2}}+
\vec{x}^{\,\prime\prime{2}})\cos(\omega\tau\sqrt{\kappa})\\
&-&2(\vec{x}^{\,\prime}\cdot{\vec{x}^{\,\prime\prime}})
\cos(\omega\tau\sqrt{\kappa-1})\nonumber
+2(\vec{x}^{\,\prime}\times{\vec{x}^{\,\prime\prime}})_{z}\sin(\omega\tau\sqrt{\kappa-1})\bigg].\nonumber
\end{eqnarray}
}
with boundary condition $\vec{x}^{\,\prime\prime}=\vec{x}(\tau)$ and $\vec{x}^{\,\prime}=\vec{x}(0)$.
Thus, one obtains the propagator (transition amplitude) as [16]
{\setlength\arraycolsep{2pt}
\begin{eqnarray}\label{9}
K_{\theta}(\vec{x}^{\,\prime\prime},\tau;\vec{x}^{\,\prime},0)
&=&\int{D[\,\vec{x}\,]}\exp\bigg(i\int^{\tau}_{0}{dt}L_{\theta}(\dot{\vec{x}},\vec{x})\bigg),\\
&=&\sqrt{det\bigg(-\frac{\partial^{2}S_{\theta}}
{\partial\vec{x}^{\,\prime\prime}\partial\vec{x}^{\,\prime}}\bigg)}e^{iS_{\theta}(\vec{x}^{\,\prime\prime},\vec{x}^{\,\prime})},\nonumber\\
&=&\frac{m\omega}{2\pi{i}\sqrt{\kappa}|\sin(\omega\tau\sqrt{\kappa})|}e^{iS_{\theta}(\vec{x}^{\,\prime\prime},\vec{x}^{\,\prime})}.\nonumber
\end{eqnarray}}
For a quantum mechanical system, the partition function can be defined in terms of the propagator of system
as [16]
\begin{equation}\label{10}
Z_{\theta}(\beta)=TrK_{\theta}(\vec{x}^{\,\prime\prime},\beta;\vec{x}^{\,\prime},0).
\end{equation}
where the inverse temperature parameter is defined as $\beta=i\tau$. The symbol $Tr$ stands for the
functional trace which for a bi-local function $A(\vec{x},\vec{x}^{\,\prime})$ in $\textit{D}$ dimensions is defined as
\begin{equation}\label{11}
TrA(\vec{x},\vec{x}^{\,\prime})=\int_{-\infty}^{+\infty}d^{D}{x}A(\vec{x},\vec{x}).
\end{equation}
When the time parameter in (11) is replaced with the inverse temperature parameter
$\beta$, the propagator (11) modifies to
\begin{equation}\label{12}
K_{\theta}(\vec{x}^{\,\prime\prime},\beta;\vec{x}^{\,\prime},0)=\frac{m\omega}{2\pi\sqrt{\kappa}\sinh(\omega\beta\sqrt{\kappa})}
e^{-S^{E}_{\theta}(\vec{x}^{\prime\prime},\vec{x}^{\prime})},\nonumber
\end{equation}
with Euclidean action
{\setlength\arraycolsep{2pt}
\begin{eqnarray}\label{13}
S^{E}_{\theta}(\vec{x}^{\,\prime\prime},\vec{x}^{\,\prime})
&=&\frac{m\omega}{2\sqrt{\kappa}\sinh(\omega\beta\sqrt\kappa)}\big[(\vec{x}^{\,\prime\,{2}}+
\vec{x}^{\,\prime\prime\,{2}})\cosh(\omega\beta\sqrt{\kappa})\\
&-&2(\vec{x}^{\,\prime}\cdot{\vec{x}^{\,\prime\prime}})
\cosh(\omega\beta\sqrt{\kappa-1})\nonumber
+2(\vec{x}^{\,\prime}\times{\vec{x}^{\,\prime\prime}})_{z}\sinh(\omega\beta\sqrt{\kappa-1})\big].\nonumber
\end{eqnarray}}
where we have invoked the well-known trigonometric identities $\cos{i\alpha}=i\cosh\alpha$ and
$\sinh\alpha=-i\sin{i}\alpha$. Thus, from the definition (13) we find for the trace of the term $e^{-S^{E}_{\theta}}$
\begin{equation}\label{14}
Tre^{-S^{E}_{\theta}(\vec{x}^{\prime\prime},\vec{x}^{\prime})}=\frac{\pi\sqrt{\kappa}}{m\omega}
\frac{\sinh(\omega\beta\sqrt{\kappa})}{\cosh(\omega\beta\sqrt{\kappa})
-\cosh(\omega\beta\sqrt{\kappa-1})}.
\end{equation}
Therefore, one is left with the partition function as
{\setlength\arraycolsep{2pt}
\begin{eqnarray}\label{15}
Z_{\theta}(\beta)&=&\frac{m\omega}{2\pi\sqrt{\kappa}\sinh(\omega\beta\sqrt{\kappa})}
Tre^{-S^{E}_{\theta}(\vec{x}^{\prime\prime},\vec{x}^{\prime})},\\\nonumber
&=&\frac{1}{4\sinh\big[\frac{\omega\beta}{2}(\sqrt{\kappa}+\sqrt{\kappa-1})\big]
\sinh\big[\frac{\omega\beta}{2}(\sqrt{\kappa}-\sqrt{\kappa-1})\big]}.
\end{eqnarray}}
Again, one recovers the familiar result holding on the commutative plane in limit $\theta\rightarrow{0}$ $(\kappa\rightarrow{1})$. The above result for the partition function is derived earlier in the context of Hamiltonian
formulation of the problem [10].
\section{Classical NC Two-Body Problem}
For the several particles living on the noncommutative plane, the $\theta$-deformed canonical structure, generalizes to
{\setlength\arraycolsep{2pt}
\begin{eqnarray}\label{16}
[\hat{x}^{A}_{i},\hat{x}^{B}_{j}]&=&i\theta_{ij}\delta^{AB},\\
{}[\hat{x}^{A}_{i},\hat{p}^{B}_{j}]&=&i\delta_{ij}\delta^{AB},\\
{}[\hat{p}^{A}_{i},\hat{p}^{B}_{j}]&=&0.
\end{eqnarray}}
where the superscripts A and B label the particles. In particular, for the two particles \textit{a} and \textit{b} bounded by a harmonic oscillator potential one finds the $\theta$-dependent Hamiltonian [11]
\begin{equation}\label{17}
H_{\theta}=\frac{1}{4m}\vec{P}^{2}_{X}+\frac{\kappa}{m}\vec{P}^{2}_{\,Y}+\frac{1}{4}m\omega^{2}\vec{Y}^{2}
+\frac{1}{2}m\omega^{2}\vec{\theta}\cdot\vec{Y}\times\vec{P}_{\,Y},
\end{equation}
where
{\setlength\arraycolsep{2pt}
\begin{eqnarray}\label{18}
\vec{X}&=&\frac{1}{2}(\vec{x}_{a}+\vec{x}_{b}),\\
\vec{Y}&=&\vec{x}_{a}-\vec{x}_{b},\\
\vec{P}_{X}&=&\vec{p}_{a}+\vec{p}_{b},\\
\vec{P}_{Y}&=&\frac{1}{2}(\vec{p}_{a}-\vec{p}_{b}).
\end{eqnarray}}
and
\begin{equation}\label{19}
[\hat{X}_{i},\hat{X}_{j}]=[\hat{Y}_{i},\hat{Y}_{j}]=2i\theta_{ij}
\end{equation}
The Hamiltonian (21) consists of a freely moving part and a $\theta$-dependent bounded term. The bounded part is same as the one-body case, Eq. (7), except that its mass and noncommutativity parameters are changed to $\frac{m}{2}$ and $2\theta$, respectively. Under these transformations the parameter $\kappa$ remains unchanged. So, contribution of the bounded part of Hamiltonian (21) to the partition function will be identical to the one made by the one-body Hamiltonian, Eq. (7). The contribution made by the free part is the familiar result of a particle of mass 2\textit{m} moving freely through a usual commutative plane. Hence
{\setlength\arraycolsep{2pt}
\begin{eqnarray}\label{20}
Z_{\theta}(\beta)&=&\int_{-\infty}^{+\infty}{d^2{p}_{a}}\int_{-\infty}^{+\infty}{d^2{p}_{b}}
\int_{-\infty}^{+\infty}{d^2{x}_{a}}\int_{-\infty}^{+\infty}{d^2{x}_{b}}
\,e^{-\frac{\beta\kappa}{2m}{(\vec{p}^{2}_{a}+\vec{p}^{2}_{b}})}\\\nonumber
&\times&\,e^{-\frac{1}{2}m\beta\omega^{2}(\vec{x}_{a}-\vec{x}_{b})^{2}}
e^{-\frac{1}{2}m\omega^{2}\vec{\theta}\cdot{}(\vec{x}_{a}\times\vec{p}_{a}+\vec{x}_{b}\times\vec{p}_{b})},\\\nonumber
&=&\int{d^{2}X}\int_{-\infty}^{+\infty}{d^{2}P_{X}}
\int_{-\infty}^{+\infty}{d^{2}P_{Y}}\int_{-\infty}^{+\infty}{d^{2}Y}
\,e^{-\frac{\kappa}{m}\beta{\vec{P}^{2}_{Y}}}\,e^{-\frac{1}{4m}\beta\vec{P}^{2}_{X}}\\\nonumber
&\times&\,e^{-\frac{1}{4}\beta{m}\omega^{2}{\vec{Y}^{2}}}
e^{-\frac{1}{2}\beta{}m\omega^{2}\vec{\theta}\cdot\vec{Y}\times\vec{P}_{\,Y}},\\\nonumber
&=&\bigg(\frac{4\pi{}m}{\beta}\bigg)\bigg(\frac{2\pi}{\beta\omega}\bigg)^{2}A.
\end{eqnarray}}
where \textit{A} stands for the area of the plane and appears after integration over \textit{X}. Again, we find a result unaffected by the noncommutativity.
\section{Quantum NC Two-Body Problem}
By following the same steps which led to the action (10), one obtains the Euclidean action of the two-body bounded system as
{\setlength\arraycolsep{2pt}
\begin{eqnarray}\label{21}
S^{E}_{\theta}(\vec{X}^{\prime\prime},\vec{Y}^{\prime\prime};\vec{X}^{\prime},\vec{Y}^{\prime})
&=&\frac{m}{\beta}(\vec{X}^{\prime\prime}-\vec{X}^{\prime})^{2}\\
&+&\frac{m\omega}{4\sqrt{\kappa}\sinh(\omega\beta\sqrt\kappa)}\big[(\vec{Y}^{\prime\,{2}}+
\vec{Y}^{\prime\prime\,{2}})\cosh(\omega\beta\sqrt{\kappa})\nonumber\\
&-&2(\vec{Y}^{\prime}\cdot{\vec{Y}^{\prime\prime}})
\cosh(\omega\beta\sqrt{\kappa-1})\nonumber\\
&+&2(\vec{Y}^{\prime}\times{\vec{Y}^{\prime\prime}})_{z}\sinh(\omega\beta\sqrt{\kappa-1})\big].\nonumber
\end{eqnarray}}
Once again, the action (28) consist of a free part and a bounded part with mass reduced to $\frac{m}{2}$ and noncommutativity parameter raised to $2\theta$. So, the imaginary time propagator takes the form
\begin{equation}\label{22}
K_{\theta}(\vec{X}^{\prime\prime},\vec{Y}^{\prime\prime},\beta;\vec{X}^{\prime},\vec{Y}^{\prime},0)\\
=\frac{m}{\pi\beta}\cdot\frac{m\omega}{4\pi\sqrt{\kappa}\sinh(\omega\beta\sqrt{\kappa})}
e^{-S^{E}(\vec{X}^{\prime\prime},\vec{X}^{\prime})}e^{-S^{E}_{\theta}(\vec{Y}^{\prime\prime},\vec{Y}^{\prime})},
\end{equation}
which from (12) and (13) leaves us with the partition function as
{\setlength\arraycolsep{2pt}
\begin{eqnarray}\label{23}
Z_{\theta}(\beta)&=&TrK_{\theta}(\vec{X}^{\prime\prime},\vec{Y}^{\prime\prime}\beta;\vec{X}^{\prime},\vec{Y}^{\prime},0),\\
&=&\frac{m}{\pi\beta}\cdot\frac{A}{4\sinh\big[\frac{\omega\beta}{2}(\sqrt{\kappa}+\sqrt{\kappa-1})\big]
\sinh\big[\frac{\omega\beta}{2}(\sqrt{\kappa}-\sqrt{\kappa-1})\big]}.\nonumber
\end{eqnarray}}
which, as expected, is just the partition function of a freely moving particle with mass 2\textit{m}, multiplied by the partition function of the one-body harmonic oscillator.
\section{Conclusions}
In this letter we derived the partition function of a classical harmonic oscillator on a noncommutative plane. We also calculated the partition function of a two-body system of particles interacting through the harmonic oscillator potential. The quantum mechanical partition function of both systems has been derived by calculating the trace of imaginary time propagators of the systems. In the $\theta\rightarrow{0}$, one recovers the familiar results on the usual commutative plane.

\end{document}